\documentclass[3p]{elsarticle}

\usepackage{amssymb}
\usepackage{amsmath}
\usepackage{amssymb}

\newcounter{bla}

\journal{Computer Physics Communications}

\usepackage{braket}
\usepackage{graphicx}

\usepackage[formats]{listings}
\usepackage{xcolor}

\lstdefinestyle{myListing}{
	numbers=left,
	stepnumber=0,
	numbersep=1pt,
	tabsize=4,
	showspaces=false,
	showstringspaces=false,
	xleftmargin=6pt,
	commentstyle=\color[RGB]{0,96,96},
	language=C
}
\usepackage{hyperref}
\hypersetup{
	colorlinks=true,
	linkcolor=blue,
	filecolor=magenta,      
	urlcolor=cyan,
}

\renewcommand{\c}[1]{\noindent\texttt{#1}}
\renewcommand{\vec}[1]{\boldsymbol{#1}}

\renewcommand{\d}[0]{\mathrm{d}}
\begin{document}

\begin{frontmatter}



\title{QEngine: An open-source C++ Library for Quantum Optimal Control of Ultracold Atoms}


\author[a]{J. J. W. H. S\o rensen}
\author[a]{J. H. M. Jensen}
\author[a]{T. Heinzel}
\author[a]{J. F. Sherson}

\cortext[author] {Corresponding author.\\\textit{E-mail address:} sherson@phys.au.dk}
\address[a]{Department of Physics and Astronomy, Aarhus University, Ny Munkegade 120, 8000 Aarhus C Denmark}

\begin{abstract}
We present the first version of the QEngine, an open-source C++ library for simulating and controlling ultracold quantum systems using optimal control theory (OCT). The most notable systems presented here are Bose-Einstein condensates, many-body systems described by Bose-Hubbard type models, and two interacting particles. These systems can all be realized experimentally using ultracold atoms in various trapping geometries including optical lattices. In addition we provide a number of optimal control algorithms including the recently introduced \textsc{group} method. The QEngine library has a strong focus on accessibility and performance. We provide several examples of how to prepare simulations of the physical systems and apply optimal control.
\end{abstract}


\end{frontmatter}



\noindent
{\bf PROGRAM SUMMARY} \\
\textit{Program Title}: QEngine \\
\textit{Program Summary}: \url{quatomic.com} \\
\textit{Download :} \url{gitlab.com/quatomic/qengine} \\
\textit{License}: MPL-2.0 \\
\textit{Programming Language}: C++14 \\
\textit{Computer}: Any system with a C++14 compliant compiler. \\
\textit{Operating system}: Linux, Mac OSX, Windows \\
\textit{RAM}: 2+ Gigabytes \\
\textit{External routines}: Armadillo, LAPACK and BLAS or Intel Math Kernel Library \\
\textit{Nature of Problem}: Quantum optimal control of ultracold systems. \\
\textit{Solution method}: Numerical simulation of the equation of motion and gradient based quantum control. \\
\textit{Running time}: Few seconds up to several hours depending on the size of the underlying Hilbert space. \\

\begin{small}	
\section{Introduction}
In the last two decades there have been exceptional advances in the ability to engineer and understand complex quantum systems. Especially, ultracold atoms provide an excellent platform for precision measurements \cite{rosi2014precision,poli2011precision}, matter wave interferometry \cite{schumm2005matter,andersson2002multimode,wang2005atom}, quantum simulation \cite{bloch2012quantum,georgescu2014quantum}, and quantum gates \cite{kaufman2015entangling,de2008optimal,weitenberg2011quantum}. These systems offer extensive versatility through their purity and the high level of control of both the underlying potential landscape and the interatomic interactions \cite{bloch2008many}. In order to fully utilize the potential of these quantum systems the design of efficient experimental protocols for preparing quantum states of interest poses an important challenge \cite{van2016optimal}.

Many experimental control protocols rely on simple empirical or adiabatic inspiration, which are typically slow and therefore limited by decoherence and sensitivity towards experimental imperfections \cite{van2016optimal,glaser2015training}. It is often desirable to find fast protocols that avoid decoherence and are robust with respect to system perturbations resulting in typically highly complex controls. Such control protocols can be found within the framework of Quantum Optimal Control (QOC). In QOC improved protocols are found using optimization algorithms that seek to minimize some cost functional \cite{werschnik2007quantum}. 

In the context of ultracold atomic physics, QOC has been applied to improve splitting and driving of Bose-Einstein condensates trapped on an atom chip, which can be used to realize matter wave interferometry and nonlinear atom optics \cite{van2016optimal,jager2013optimal,jager2014optimal}. QOC has also been applied to stabilize ultracold molecules \cite{ndong2010vibrational,koch2004stabilization} and manipulate ultracold many-body systems in optical lattices \cite{doria2011optimal}. In addition, it has been demonstrated that such optimal control pulses are experimentally feasible \cite{van2016optimal,bucker2013vibrational}. There has also been fundamental studies showing that QOC can find controls saturating the fundamental quantum speed limit \cite{caneva2009optimal,hegerfeldt2013driving,deffner2017quantum} where similar behavior has also been reported for ultracold atoms in a double well system \cite{brouzos2015quantum}. 
QOC is a versatile tool that can be applied not only in the context of ultracold atoms but as examples also in nuclear magnetic resonance \cite{khaneja2005optimal}, control of chemical reactions \cite{assion1998control} and nitrogen vacancies \cite{dolde2014high}. 

Optimal control protocols are typically designed for a certain set of experimental parameter values that may change due to modifications in the experimental setup, thereby necessitating a recalculation for new optimal controls. For instance, there is a large number of papers that discuss driving a condensate from the ground state into the first excited state with slightly different parameter values \cite{van2016optimal,jager2013optimal,jager2014optimal,bucker2013vibrational,mennemann2015optimal,hohenester2007optimal}.
A barrier for rapidly recalculating such controls is writing and rigorously testing QOC programs, which is time consuming and the programs are very slow if not properly implemented. These two requirements, performance and usability, are primary driving forces behind the design of the QEngine. 
The high level of usability for example enables experimentalists to readily recalculate experimental protocols.

There exist a number of alternative software packages to the QEngine for performing quantum optimal control. Many of these are implemented in \textsc{matlab} like \textsc{octbec} \cite{hohenester2014octbec}, \textsc{dynamo} \cite{machnes2011comparing} and the recent WavePacket \cite{schmidt2017wavepacket1,schmidt2018wavepacket}. The Python package QuTiP is also a widespread platform for simulation and optimal control of quantum optics \cite{johansson2012qutip}. Collectively these packages offer more functionality than the QEngine but they are implemented in weakly typed programming languages that are inherently less focused on performance. Especially, \textsc{octbec} has a similar focus to the QEngine and it has been a source of inspiration for our work.

The QEngine is designed for performance. One way the QEngine achieves this is by a general reliance on templates to provide flexibility instead of virtual functions and pointers. This allows for a high number of compile-time optimizations.
Templates are useful for efficiency, but they are typically a programming barrier for physicists who are not C++ professionals. In order to accommodate such users of the library, we have made considerable efforts in providing a simplified API that does not invoke any advanced language features. The \c{auto}-syntax available in modern C++ together with factory-functions and overloaded operators give a straightforward syntax close to the mathematical equations used by theoretical physicists and weakly typed languages such as \textsc{matlab} and Python. 
The QEngine uses the highly optimized Intel Math Kernel Library (MKL) and the C++ library Armadillo to provide efficient basic linear algebra needed for the quantum simulations and optimizations \cite{armadillo}. In addition, the code has also been profiled and optimized. 

A comprehensive documentation for the QEngine is also available at \url{quatomic.com} and the source code is available at \url{gitlab.com/quatomic/qengine}. The library has a number of example programs that can help users get started. In this paper, we give an introduction to some of the features in QEngine but leave out several details that can be found in the online documentation. The
QEngine currently supports simulation and optimal control of the Gross-Pitaevskii description of a BEC, the Bose-Hubbard model, two interacting particles, a single particle, and generic few mode models.

The paper is organized as follows. In section \ref{sec:Overview} we give a brief introduction to the physical models and optimal control theory. In section \ref{sec:simulation} we discuss how to prepare simulations in two example programs that demonstrate key functionalities in the QEngine. Quantum optimal control theory is explained in section \ref{sec:qoc} including the \textsc{group} algorithm we recently introduced in Ref. \cite{sorensen2018group}. Finally in section \ref{sec:qocPrograms} we explain how to perform optimal control optimizations on the example programs from section \ref{sec:simulation}. Section \ref{sec:summaryOutlook} gives a summary and outlook.
\begin{figure}[t]
	\centering
	\includegraphics[width=\linewidth]{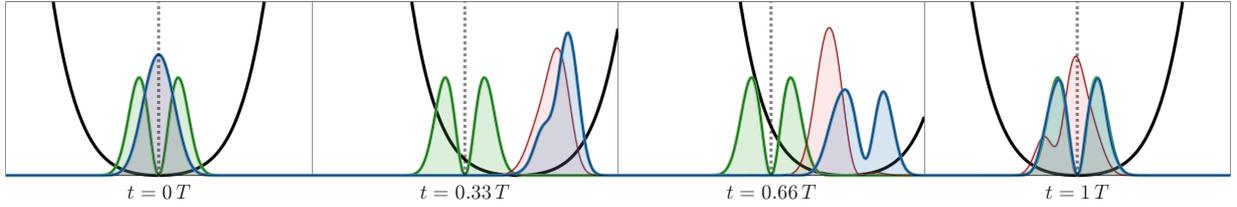}
	\caption{(Color Online) Snapshots of an optimized (blue)  and unoptimized (red) transfer of a ground state BEC into the first excited state (green) in an anharmonic potential (black) -- see section \ref{subsec:gpeex} for details. The simulations and optimization were performed using the QEngine.} 
	\label{fig:shakeup}
\end{figure}

\section{Overview of QEngine Features} \label{sec:Overview}
In this section we give a brief overview of the mathematical description of the models available in the QEngine and optimal control. The starting point for modelling these ultracold atomic systems is the second quantized Hamiltonian \cite{dalfovo1999theory}
\begin{align}
\hat H = \int \left(\hat\Psi^\dagger (x) \left[-\frac{\hbar^2}{2m} \frac{\partial^2}{\partial x^2 }+ \hat V \right] \hat \Psi(x) + \frac{g_{\text{1D}}}{2} \hat \Psi^\dagger(x) \hat \Psi^\dagger(x) \hat \Psi(x)\hat \Psi(x)\right) \d x.
\label{eq:hamiltonian}
\end{align}
Here the first term describes the kinetic and potential energy and the second term represents two-particle interactions. Currently, the QEngine only supports one-dimensional systems. $\hat \Psi(x)$ and $\hat \Psi^\dagger(x)$ are the bosonic field operators obeying the usual commutation relations $[\hat \Psi(x_1), \hat \Psi^\dagger(x_2)]=\delta(x_1-x_2)$.
$\hbar$ is the reduced Planck constant and $m$ is the mass of the atom. 
In Eq. \eqref{eq:hamiltonian} we have used that  two-particle interactions in the ultracold limit are well-described by an effective contact interaction, which in one-dimension is $V_{\text{int}}(x_1,x_2) = g_{\text{1D}} \delta(x_1-x_2)$ where $g_{\text{1D}}$ is the system dependent coupling strength \cite{bloch2008many,olshanii}.
Different physical models described in the QEngine emerge from different special cases of Eq. \eqref{eq:hamiltonian}.

\paragraph{Gross-Pitaevskii Equation}
One important special case for $N$ ultracold bosons is a mean field description where the bosonic field is in a single mode. This gives rise to the Gross-Pitaveskii Equation (GPE) describing the time evolution of a Bose-Einstein condensate (BEC) $\psi = \sqrt{N} \phi$. The corresponding Hamiltonian is
\begin{gather}
\hat H_{\text{gp}}= \hat H_0 + g_{\text{1D}} |\psi(x,t)|^2,
\label{eq:GPE}
\end{gather}
where $\hat H_0$ is the kinetic and potential energy. The non-linear term represents the condensate self-interaction. The GPE is an important starting point for modelling the dynamics of BECs \cite{schumm2005matter,bloch2008many,van2016optimal,jager2013optimal}. The QEngine can calculate the ground state and first excited state of the GPE using an optimal damping algorithm described in Ref. \cite{dion2007ground}. The time evolution is performed using the split-step Fourier method \cite{taha1984analytical}.

\paragraph{Bose-Hubbard}
Ultracold atoms can be loaded into a periodic optical lattice \cite{bloch2008many}. In this system it is convenient to expand $\hat \Psi(x)$ in terms of the localized Wannier modes on each lattice site. In a lowest band approximation the expansion reads $\hat \Psi(x) = \sum_i \hat a_i w_0(x-x_i)$. 
Also assuming the tight-binding approximation Eq. \eqref{eq:hamiltonian} for $L$ lattice sites becomes
\begin{align}
\hat H = -J\sum_{i=1}^{L-1} \left(\hat a_{i+1}^\dagger \hat a_{i} + \text{h.c}.\right) + \frac{U}{2}\sum_{i=1}^{L} \hat n_i(\hat n_i-1) + \sum_{i=1}^L V_i \hat{n}_i \label{eq:BH}
\end{align}
Here $J$ and $U$ are matrix elements of Eq. \eqref{eq:hamiltonian} with the lowest band Wannier functions, which describe the nearest-neighbor tunneling and on-site interaction. $V_i$ is the local external trapping potential.
The ground state of this Hamiltonian exhibits a phase transition from a delocalized superfluid state to a Mott insulating state depending on the ratio $U/J$ \cite{bloch2008many}. This model is simulated in the QEngine using exact diagonalization with sparse linear algebra. The time evolution is performed using the Lanczos method \cite{park1986unitary}.

\paragraph{Two-particle}
Technological advances have enabled the preparation of single atoms in an optical lattice or tweezer arrays \cite{bloch2008many,barredo2016atom,endres2016atom}. It has been proposed to use these systems as a platform for quantum computation, where the necessary two-qubit gate can be realized using controlled ultracold collisions of two atoms \cite{kaufman2015entangling,weitenberg2011quantum,deChiaraGate,anderlini2007controlled}.  
It is convenient to rewrite Eq. (\ref{eq:hamiltonian}) in first quantization as  
\begin{equation}
\hat H = -\frac{\hbar^2}{2m}\frac{\partial^2}{\partial x_1^2} +\hat V(x_1) -\frac{\hbar^2}{2m}\frac{\partial^2}{\partial x_2^2}+\hat V(x_2) + g_{\text{1D}}\delta(x_1-x_2), \label{eq:TwoParticle}
\end{equation}
where $x_1$ and $x_2$ are the positions of the two atoms. The associated dynamics is also simulated using the split-step Fourier method. In a similar manner it is possible to simulate the dynamics of a single particle in the QEngine.

\paragraph{Units}
In order to perform any physical simulation it is convenient to transform the Hamiltonians in Eqs. (\ref{eq:GPE})-(\ref{eq:TwoParticle}) into dimensionless units. 
A discussion of the units used in the example programs is given in the appendix.

\paragraph{Quantum Optimal Control} The QEngine enables the user to solve state transfer problems using QOC. 
This type of problem consists in manipulating the system dynamics in order to realize a transfer of an initial state $\psi_0$ into a target state $\psi_t$ for some fixed duration $T$. The manipulatory access to the dynamics is through one or more control fields $u(t)$ parametrizing the Hamiltonian in some way $H=H(u(t))$. In an experimental setting
the control fields usually correspond to physical quantities such as the intensity or position of a laser beam.
Optimization algorithms are typically used to iteratively design the control fields.  
An example of a state transfer problem is shown in Fig. \ref{fig:shakeup} where a BEC is driven from the ground state into the first excited state using an optimized control field. In this case the control field corresponds to the position of the trap center, which is experimentally realized by adjusting magnetic fields \cite{van2016optimal,bucker2013vibrational}. The figure shows snapshots of the transfer process before and after the optimization, illustrating that the optimization algorithm succeeds in finding an optimal control. 

The QEngine offers a variety of optimization algorithms that can be applied to any of the physical models. 

\paragraph{Benchmark against \textsc{matlab}}
For comparison we benchmarked the QEngine against a similar implementation in \textsc{matlab} used internally in our research group. The benchmark was performed on the entire Gross-Pitaevskii example program described below, which ran for 100 optimization iterations with different grid sizes. The example program was slightly modified to give the most direct comparison between the two code bases. The results are displayed in Fig. \ref{fig:benchmark}, which shows that the QEngine is significantly faster.
\begin{figure}[t]
	\centering
	\includegraphics[width=0.65\linewidth]{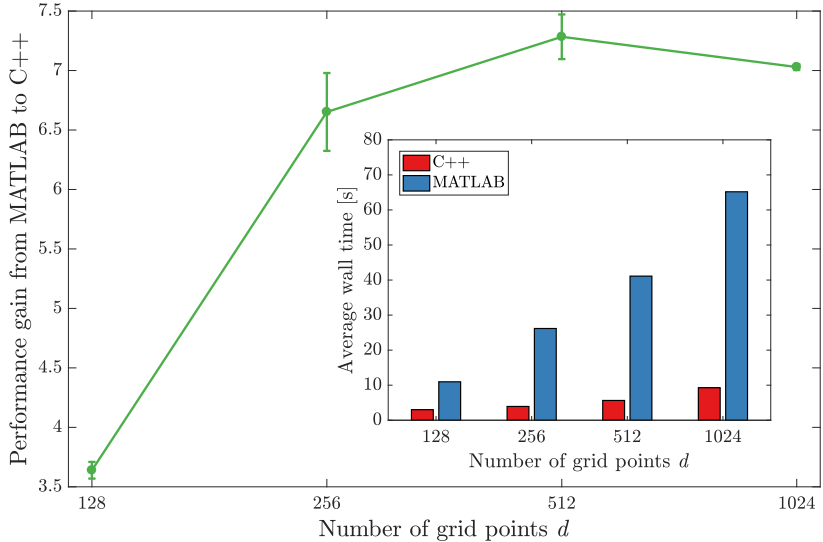}
		\caption{The relative performance gain from \textsc{matlab} to C++ (higher is better) when running 100 optimization iterations in the Gross-Pitaevskii example program. The results are averaged over 15 individual runs. The inset shows the average absolute execution time for \textsc{matlab} and C++ (lower is better). The results were generated on a 2017 Macbook Pro laptop with an Intel(R) Core\texttrademark\;i7-7700HQ CPU @ 2.80GHz processor.}
		\label{fig:benchmark}
\end{figure}

\section{Simulation Example Programs} \label{sec:simulation}
Instructions for installing the QEngine are included in the \c{README.md} file. In order to help users get started we have included a number of example programs in the folder \c{example\_projects} that are designed to illustrate many of the different features in QEngine. An overview of these programs is given in Table \ref{tbl:programs}.
In this paper, we discuss the two programs \c{gpe-example.cpp} and \c{bosehubbard-example.cpp} in depth, which gives a good overview of key functionalities in the QEngine. For clarity, we omit minor code details in this paper.
A more detailed and complete API documentation can be found at \url{quatomic.com}, including more technical functionalities not discussed in the example programs.

\begin{table}[t]
\begin{center}
\begin{tabular}{l c l}
\hline
Example Program & Execution Time & Description \\
\hline
gpe-example.cpp & 582.27s & Optimal control of driving a condensate wave function. \\
bosehubbard-example.cpp & 234.94s & Optimal control of a superfluid to Mott transfer. \\
twoparticle-example.cpp & 975.32s & Optimal control of a an ultracold atomic gate. \\
oneparticle-example.cpp & 0.99s & Optimal control of a single atom in a optical tweezer. \\
twolevel-example-cpp & 0.67s & Optimal control of the Landau-Zener system. \\
\hline
\end{tabular}
\caption{The default example programs included in the QEngine. The table shows the file name, execution time and a short description. The results were generated on a 2017 Macbook Pro laptop with an Intel(R) Core\texttrademark\;i7-7700HQ CPU @ 2.80GHz processor.}
\label{tbl:programs}
\end{center}
\end{table}

\subsection{Gross-Pitaevskii Example Program}
\label{subsec:gpeex}
As a first example, we describe optimal control of a BEC trapped on an atom-chip. 
We focus on the control problem described in Refs. \cite{van2016optimal,jager2014optimal,bucker2011twin} where a BEC is transferred from the ground state into the first excited state as shown in Fig. \ref{fig:shakeup}.                                                            
The physical motivation for this problem is to create a source of twin-atom beams, which is the matter-wave analogue to twin-photon beams \cite{bucker2011twin}.
The mechanism behind  twin beam emission is binary collisions of two excited atoms. The collision may cause atoms to de-excite into the radial ground state mode while simultaneously populating twin momentum states $\ket{\pm k_0}$ along the axial $z$-direction due to conservation of momentum and energy. 
The characteristic timescale for the collision induced decay is a few milliseconds ($\approx3$ms), and as a consequence the duration of the preparation stage into the excited state must be well below this threshold \cite{van2016optimal}. 
The atom-chip experiment has two tightly confined transverse directions ($x$ and $y$) and a weakly confined axial direction ($z$). One of the transverse directions (say $y$) has a tighter confinement freezing out excitations. The dynamics along the axial direction is slow compared to the transverse directions, and we may only consider a one-dimensional GPE along the $x$-direction. This requires an appropriate effective coupling constant \cite{van2016optimal,gerbier2004quasi}. 

The potential along the $x$-direction is parameterized by a single control field $u(t)$, which is well-approximated by the anharmonic potential
\begin{align}
V(x,u(t)) = p_2(x-u(t))^2 + p_4(x-u(t))^4 + p_6(x-u(t))^6,
\label{eq:shakeup}
\end{align}
where the $p_i$'s are constants obtained experimentially \cite{van2016optimal,bucker2011twin}. 
The initial state is taken to be the ground state of $V(x,0)$, and in the optimal control part we take the target state to be the first excited state of $V(x,0)$ as shown in Fig. \ref{fig:shakeup}. Measuring length in units of micrometers and time in units of milliseconds, the effective mean field interaction strength for 700 atoms is $g_{\text{1D}}=1.8299$ 
(see \ref{ap:Units}). 

To use the QEngine in a program, we need to include the QEngine header file 
\begin{lstlisting}
#include <qengine/qengine.h>
#include <iostream>

using namespace qengine;
\end{lstlisting}
The \c{qengine.h} header exposes the different simulation models and optimal control algorithms.
At the highest level the QEngine library defines the namespace \c{qengine}. The \c{qengine} namespace contains most API-functionality across the different types of physics and optimal control. 
We will make use of the \c{DataContainer} class defined in the QEngine, which can be used to save data to a .json file format or optionally to a \textsc{matlab} .mat file format. This makes it easy to export data for visualization and post-processing.

First we set up the control field $u(t)$ 
\begin{lstlisting}
const auto dt = 0.002;
const auto duration = 1.25; // corresponds to 1.25ms
const auto n_steps = floor(duration/dt) + 1;

const auto ts = makeTimeControl(n_steps,dt);
const auto initialAmplitude = 0.55;
const auto u = initialAmplitude*sin(PI/duration*ts); // control field
\end{lstlisting}

The \c{makeTimeControl} function returns a single control field with linearly spaced values, which can be used to compose more complicated control fields. In this case the control field \c{u} is half a sine period with amplitude 0.55. This will also act as our initial guess in the optimal control algorithms.
It is possible to access the control field values at time index $i$ by calling \c{u.get(i)}, returning an \c{RVec} whose entries are the values for each control field at that time index. The control field values at the first and last time index can be easily accessed with \c{u.getFront()} and \c{u.getBack()}, respectively. In the present case we only have a single control field. 

The concept of a Hilbert space is mimicked in the QEngine for each type of physical model. 

\begin{lstlisting}
const auto kinFactor = 0.36537;  // T = -kinFactor*d^2/dx^2
const auto s = gpe::makeHilbertSpace(-2,+2,256,kinFactor);
const auto x = s.x(); // FunctionOfX of x-grid values
\end{lstlisting}

Having defined both $x$ and $u(t)$ we can create the control-dependent potential $V(x,u(t))$  Eq. \eqref{eq:shakeup} 
\begin{lstlisting}
const auto p2 =  65.8392;
const auto p4 =  97.6349;
const auto p6 = -15.3850;

const auto V_func = [&x,p2,p4,p6](const real u)
{
     // By saving intermediate calculations we reduce overall computation time
	const auto x_u = x - u;
	const auto x_uPow2 = x_u * x_u;
	const auto x_uPow4 = x_uPow2 * x_uPow2;
	const auto x_uPow6 = x_uPow2 * x_uPow4;
	
	return  p2*x_uPow2 + p4*x_uPow4 + p6*x_uPow6;
};

const auto u_initial = u.getFront().front(); // first entry in first time index 
const auto V = makePotentialFunction(V_func, u_initial);
\end{lstlisting}
The lambda function \c{V\_func} takes a \c{real} number
 and returns a \c{FunctionOfX} evaluated with the given control value. 
To create a potential object the lambda function and an initial control field value are combined in  \c{makePotentialFunction}.
The \c V object encapsulates the idea of a potential, and calling \c{V(newControlValue)} evaluates 
the \c{V\_func} lambda with \c{newControlValue}
and returns a potential operator. 
In the present case \c{newControlValue} is of type \c{real}.
The kinetic energy operator can simply be extracted from the Hilbert space. It is represented by the 5-diagonal approximation to the second derivative with non-periodic boundary conditions.
The mean field interaction is equally succinctly handled. Assembling the Hamiltonian operator is then straightforward,
\begin{lstlisting}
const auto T = s.T(); 
const auto g1D = 1.8299;
const auto meanfield = makeGpeTerm(g1D);

const auto H = T + V + meanfield; 
\end{lstlisting}
The \c{H} object can be called in the same way as the underlying potential by \c{H(newControlValue)}. Note that the type of \c{H} is \c{auto} deduced to be a GPE Hamiltonian by the compiler. Omitting the \c{meanfield} term would change the type deduction to a single particle Hamiltonian. 
This would still valid code since the GPE Hilbert space is the same as the single particle Hilbert space.
The QEngine defines a convenient syntax for creating general linear combinations of eigenstates for operators. Let \c{A} be an operator and let $\{\phi_i\}$ be the corresponding eigenstates and suppose we wanted to create the linear combination $\psi = \phi_0 - 2i \phi_1$. This is readily achieved with the lines,
\begin{lstlisting}
const auto comb = A[0] - 2.0*1i*A[1];  	  // syntax object
const auto psi = makeWavefunction(comb);  // evaluate syntax 
\end{lstlisting}
The states we need for the example are individual eigenstates
\begin{lstlisting}
const auto psi_0 = makeWavefunction(H(u.getFront())[0]);
const auto psi_t = makeWavefunction(H(u.getBack() )[1]);
\end{lstlisting}
The QEngine currently only supports calculation of the ground state and first excited state for GPE type physics, but will not raise an error if higher excited states are queried. There are no such restrictions for the other types of physics.

Initializing the container and filling in time independent data is done by
\begin{lstlisting}
DataContainer dc; // empty data container
dc["dt"] = dt;
dc["duration"] = duration;
dc["x"] =  x.vec();
dc["psi_t"] = psi_t.vec();
\end{lstlisting}
Once all data has been collected, calling \c{dc.save("<path/to/dest>.<file-extension>")} 
creates a file in either the .json or .mat file format, where e.g. the variable \c x is stored with the field name "\c x" and corresponding field values. 

To perform time evolution we initialize a stepper with fixed stepping interval length \c{dt}. We then loop over the entire control and append a few quantities of interest at each instant of time for saving.
\begin{lstlisting}
auto stepper = makeFixedTimeStepper(H,psi_0,dt);
for(auto i = 0; i < n_steps; i++)
{
	const auto& psi = stepper.state();
	dc["V"].append(V(u.get(i)).vec());
	dc["psis"].append(psi.vec());
	dc["overlap"].append( overlap(psi, psi_t));
	dc["fidelity"].append(fidelity(psi,psi_t));
	dc["x_expect"].append(expectationValue(x,psi));
	if(i < n_steps-1) stepper.step(u.get(i+1));
}
\end{lstlisting}
The propagation from $\psi(t)$ to $\psi(t+dt)$ is performed by \c{stepper.step(u.get(i+1))} using the midpoint rule. Only $u(t+dt)$ is needed since $u(t)$ is stored internally from the previous step.

In the program we subsequently take additional steps with the final Hamiltonian held constant by using the \c{stepper.cstep()} function in an otherwise identical loop to the one above.

The result of the simulation is illustrated in Fig. \ref{fig:gpBefore} where the density of the condensate is plotted as a function of time. In the atom-chip experiment the objective is to transfer the initial state into the first excited state, which is not accomplished in the unoptimized transfer Fig. \ref{fig:gpBefore}.

\subsection{Bose-Hubbard Example Program}
\label{subsec:bosehubbard}
In this example, we discuss the simulation of bosons in an optical lattice described by the Bose-Hubbard model Eq. \eqref{eq:BH}. Creating a Mott state with one particle on each site is important for many experimental applications such as quantum logic gate operations \cite{two_qubit_quantum_gate_by_cold_controlled_collisons, quantum_gate_BH_via_collisions,laser_induced_quantum_gate_operations,FastQuantumGates,Toffili_gate_1D_lattice}, quantum simulation \cite{quantum_simulator}, and single atom transistors \cite{single_atom_transistor}. 
Experimentally the system is initialized in the superfluid state and must be dynamically transferred into the Mott state  \cite{braun2015emergence, BH_experimental_realization}. However, near the phase transition the gap between the ground state and the first excited state closes in an infinite system. This implies diverging transfer times for adiabatic solutions. There have been both experimental and numerical attempts to find improved transfer protocols \cite{doria2011optimal,braun2015emergence,ramping_optimization_zakrzewski,rosi2013fast}.

Here we consider a transfer from the ground state at $U=4$ into a Mott like ground state at $U=30$ with a weak harmonic external potential in units of \textit{J}. It is necessary to impose a minimal $U_{\text{min}} = 2$ since the Bose-Hubbard model assumes a sufficiently deep lattice \cite{bloch2008many}. 
In a similar manner, it is not experimentally feasible to have arbitrarily large values of $U$. We take the upper bound to be $U_{\text{max}}=40$.
Later we will apply QOC to find optimized solutions, which must also satisfy these experimental and modelling constraints on $U$. The constraints can be accommodated by introducing a nonlinear transformation $U(u) = A(\tanh(u) + B)$, where $u$ is a non-physical but unbounded control field. Here $A=U_{\text{max}}/(1+B)$ and $B = (1+U_{\text{min}}/U_{\text{max}})/(1-U_{\text{min}}/U_{\text{max}})$ restricts the physical control $U_{\text{min}} <  U < U_{\text{max}}$.

\begin{lstlisting}
const auto Umin = 2.0;
const auto Umax = 40.0;

const auto B = (1+Umin/Umax)/(1-Umin/Umax); // transformation params
const auto A = Umax/(1+B);

const auto UFromu = [A,B](auto u){return A*(tanh(u) + B);}; // control to physical U
const auto uFromU = [A,B](auto U){return atanh(U/A-B);};    // physical U to control

const auto dt = 0.002;
const auto duration = 2.2; 
const auto n_steps = floor(duration / dt) + 1;

const auto ts = makeTimeControl(n_steps, dt);
const auto u = uFromU(Umin + 0.5*exp(log((30-Umin)/0.5)*ts/duration));
\end{lstlisting}
Here we use an exponential ramp for $U$, which will later be used as the starting point for the QOC algorithms \cite{doria2011optimal}.  As in the previous example we begin by creating the underlying Hilbert space and subsequently initialize the terms in the Hamiltonian. We also demonstrate how to add a weak confinement potential.

\begin{lstlisting}
const auto space = bosehubbard::makeHilbertSpace(5,5);

const auto periodicBoundaries = false;
const auto hoppingOperator = space.makeHoppingOperator(periodicBoundaries);
const auto onSiteOperator  = space.makeOnSiteOperator();

const auto sitePositions = linspace(-1.0, +1.0, space.nSites());
const auto potential = 0.1*pow(sitePositions,2);
const auto V = space.transformPotential(potential); // transform to site indices

const auto H_J = -1.0* hoppingOperator; // J = 1.0
const auto H_const = H_J + V; // Constant parts of Hamiltonian is added

const auto H_func = [&H_const, &onSiteOperator, &UFromu](const real u)
{
	return H_const + 0.5*UFromu(u)* onSiteOperator;
};
	
const auto H = makeOperatorFunction(H_func,u.getFront().front());
\end{lstlisting}
As in the previous example the full Hamiltonian \c{H} is assembled using a lambda function \c{H\_func} and an initial control value \c{u.getFront().front()}.
After initializing the Hamiltonian we set up the superfluid state and the Mott like state. We then initialize the time stepper. The default stepper is a Lanczos propagator, which uses a user supplied Krylov order \cite{park1986unitary}. 
\begin{lstlisting}
const auto psi_0 = makeState(H(u.getFront())[0]);
const auto psi_t = makeState(H(u.getBack())[0]);

const auto krylovOrder = 4;
auto stepper = makeFixedTimeStepper(H,psi_0,krylovOrder,dt);
\end{lstlisting}
These lines of code complete the necessary steps to set up a Bose-Hubbard simulation. 
Exactly as in the previous example we propagate over the control and in this case save the single-particle density matrix by
\begin{lstlisting}
dc["rho1"].append(space.singleParticleDensityMatrix(state));
\end{lstlisting}
\c{state} is the instantaneous state from the \c{stepper} when propagated over the control \c{u}. The result of the Bose-Hubbard simulation is illustrated in Fig. \ref{fig:bhBefore} where the on-site density is plotted as a function of time. In the superfluid-Mott transfer the objective is to reach the Mott insulator type state, which is not accomplished in the unoptimized transfer Fig. \ref{fig:bhBefore}.

\begin{figure}[t]
\begin{minipage}{0.48\linewidth}
	\centering
	\includegraphics[width=1\linewidth]{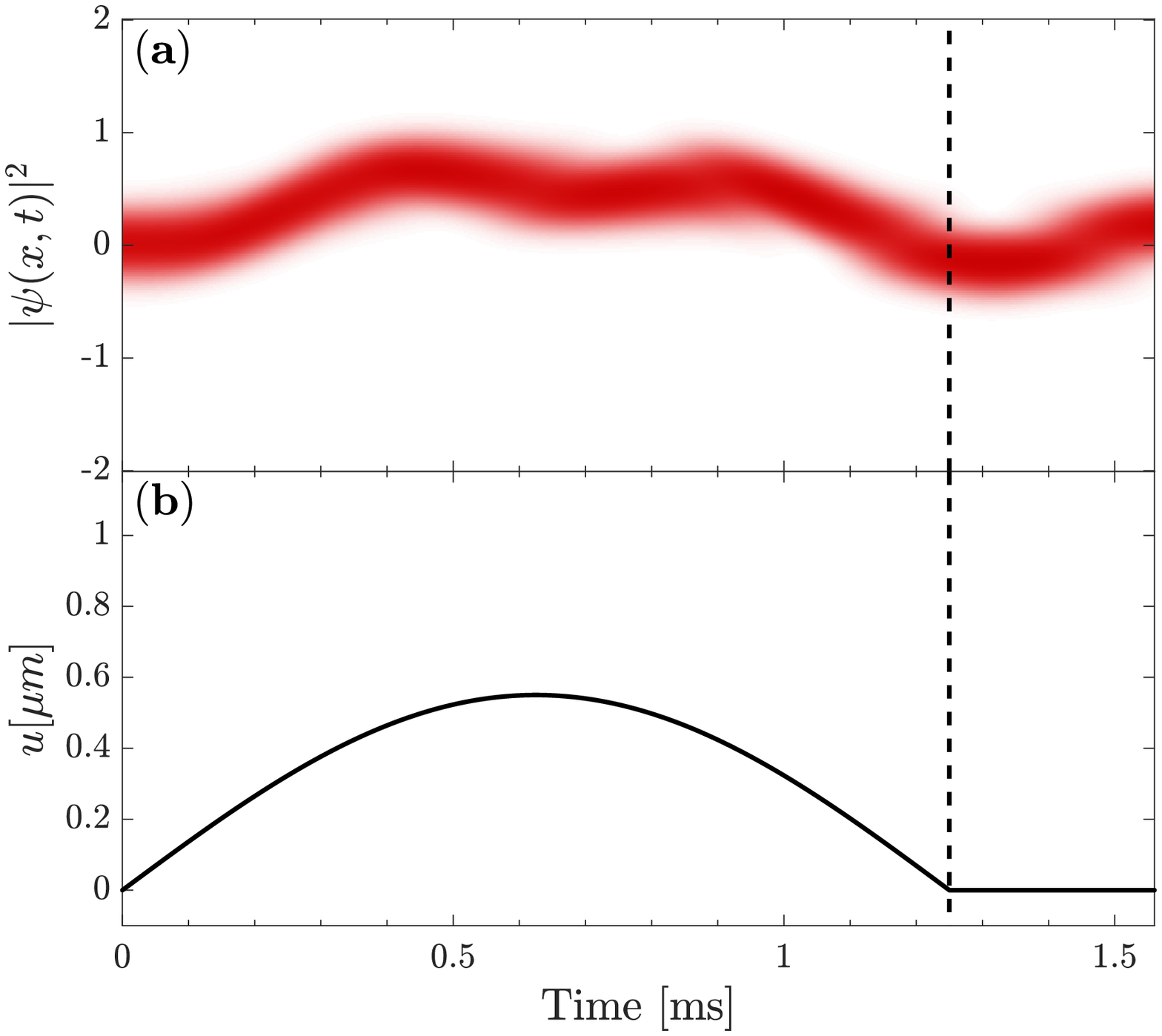}
		\caption{(\textbf{a}) the density of the condensate $|\psi(x,t)|^2$ when propagated along the unoptimized control (\textbf{b}) from the Gross-Pitaevskii example program. The initial control gives $F=0.23$. After the vertical dashed line the control is held constant.}
		\label{fig:gpBefore}
\end{minipage}
\quad
\begin{minipage}{0.48\linewidth}
		\centering
	\includegraphics[width=1.07\linewidth]{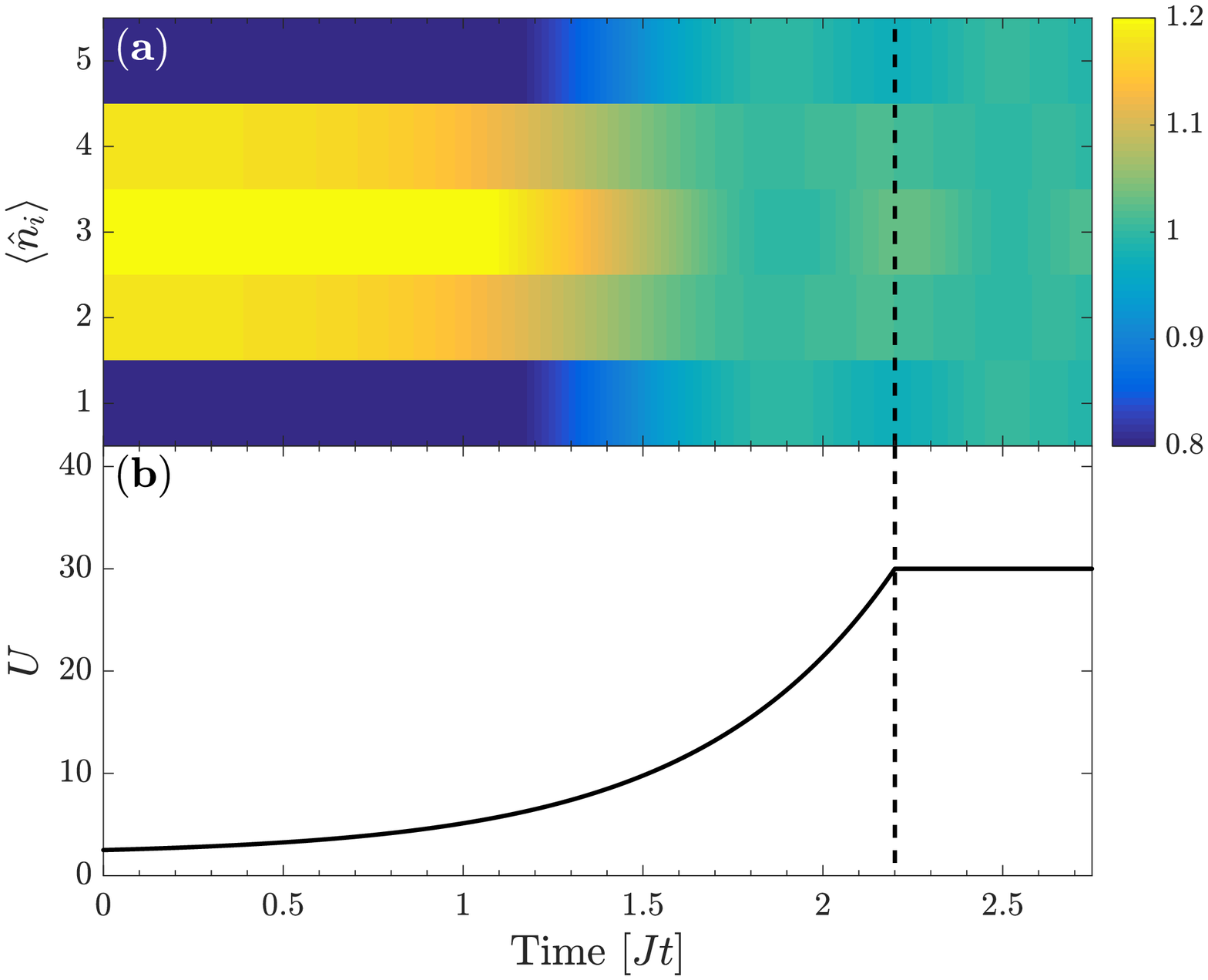}
	\caption{(\textbf{a}) the on-site density $\braket{\hat n_i}$ when propagated along the unoptimized control (\textbf{b}) from the Bose-Hubbard example program. The initial control gives $F=0.81$. After the vertical dashed line the control is held constant.}
	\label{fig:bhBefore}
\end{minipage}

\end{figure}

\section{Theory of Quantum Optimal Control} \label{sec:qoc}
In the previous section we have described how to set up simulations in the QEngine. In this section we briefly review QOC and related algorithms. In the next section we describe how to apply QOC to the example programs. 

Consider the problem of engineering a single control field $u(t)$ realizing the state transfer from $\psi_0$ to $\psi_t$ in duration $T$ constrained by the equation of motion $i\hbar\dot\psi = \hat H(u)\psi$ for all $t$. We may consider $\hat H=\hat H_{\text{gp}}$ to be the general case, as taking $\beta=0$ produces the usual Schr\"{o}dinger equation. 
In QOC this problem is posed as a minimization of the cost functional \cite{sorensen2018group}
\begin{align}
J[\psi,\chi,u] &= J_\mathcal{F}[\psi] + J_\gamma[u] + J_{\text{gp}} [\psi,\chi,u] \\
&= \frac{1}{2}\left(1-|\braket{\psi_t | \psi(T)}|^2\right) + \frac{\gamma}{2}\int_0^T\dot u^2 \d t + \Re\int_0^T\Bigl\langle \chi \Bigl| \Big(i\hbar\partial_t - \hat H_0(u) - \beta |\psi|^2 \Big) \Bigr|\psi\Bigr\rangle \d t, \label{eq:cost}
\end{align}
where the time dependence of most quantities has been suppressed for readability.
The first term is minimal for perfect transfers up to a global phase i.e. when the fidelity $F=|\braket{\psi_t|\psi(T)}|^2$ is 1. The second term penalizes rapid temporal fluctuations in the control field, which are typically not experimentally feasible.  The relative importance between the first and second term is determined by a regularization hyperparameter $\gamma \geq 0$ where higher values shifts preference towards smoother controls. Usually $\gamma \sim 10^{-7}-10^{-5}$. The last term containing the Lagrange multiplier $\chi(t)$ ensures the equation of motion is obeyed at all times.

\subsection{\textsc{grape}}
Setting the first G\^ ateaux variations of $J$ wrt. the functions $\{\psi(t),\chi(t),u(t)\}$ to zero 
\begin{align}
D_{\delta\psi}J = D_{\delta\chi}J = D_{\delta u}J = 0,
\end{align}
and assuming the variations of the control vanish at the boundaries ($t=0$ and $t=T$) lead to the first order optimality conditions \cite{sorensen2018group}
\begin{align}
i\hbar{\dot\psi} &= (\hat H_0(u) +\beta |\psi|^2)\psi, & \psi(0) &= \psi_0, \label{eq:optimality1}\\
i\hbar{\dot\chi} &= \left(\hat H_0(u)  +2\beta|\psi|^2 \right) {\chi}+\beta\psi^{2} {\chi^*},& {\chi(T)} &=i \braket{\psi_t|\psi(T)}{\psi_t},\label{eq:optimality2}\\
\gamma \ddot u &= -\Re\Braket{\chi | \frac{\d \hat H_0(u)}{\d u} |\psi}, & u(0)&=u_0,\quad u(T)=u_T. \label{eq:optimality3}
\end{align}
At this point we may think of $J$ as a functional of only $u$, $J=J[u]$, with the corresponding dynamics of $\psi$ and $\chi$ determined by the equations of motion above. An analytical approach to solving this set of equations is not generally feasible. However, we may define the gradient of $J$ wrt. $u(t)$ under the $X$ norm as the element $\nabla_X J$ fulfilling the relation 
\begin{align}
D_{\delta u} J = \langle \nabla_X J, \delta u \rangle_X , \label{eq:gradientdef}
\end{align}
where $\delta u$ is an arbitrary variation.
The common choices of the norm are $X=L^2$ and $X=H^1$ defined as $\langle f, g \rangle_{L^2} = \int_0^T f(t) g(t) dt$ and $\langle f, g \rangle_{H^1} = \int_0^T \dot f(t) \dot g(t) dt$, respectively \cite{von2008computational} . 
Eq. \eqref{eq:gradientdef} establishes an indirect way of calculating the gradient and for the norms above we obtain

\begin{align}
\nabla_{L^2}J(t) &=  - \Re\Braket{\chi(t) | \frac{\d H_0(u(t))}{\d u} | \psi(t)} -\gamma \ddot{u}(t) \label{eq:L2},\\
{\nabla_{H^1}\ddot J}(t) &= - \nabla_{L^2}J(t) \label{eq:H1} .
\end{align}
The $L^2$ gradient may not vanish at the boundaries so we must artificially enforce $\nabla_{L^2}J(0) = \nabla_{L^2}J(T) = 0$ to respect Eq. \eqref{eq:optimality3}. 
In solving the Poisson Eq. \eqref{eq:H1} for the $H^1$ gradient we may conveniently choose Dirichlet boundary conditions $\nabla_{H^1}J(0) = \nabla_{H^1}J(T) = 0$ directly so Eq.  \eqref{eq:optimality3} is always fulfilled.

The control is iterated towards a local minimum of the cost functional by the update rule
\begin{align}
u^{(k+1)} = u^{(k)} + \alpha^{(k)} p^{(k)},
\label{eq:grapeupdate}
\end{align}
where $\alpha^{(k)}$  is a suitable step size along the search direction $p^{(k)}$ for the $k$'th iteration. Typically the search direction is based on gradient information \cite{nocedal2006numerical}. The simplest choice is searching in the direction of steepest descent $p_{\text{SD}}^{(k)} = - \nabla_{X}J[u^{(k)}]$ where we again are free to choose either $X=L^2$ or $X=H^1$.  
Another common search direction is the Newton direction $p_{\text N}^{(k)} = (\nabla_X^2 J[u^{(k)}])^{-1} \nabla_X J[u^{(k)}]$, which takes into account the local curvature of the functional. This requires an expensive calculation of the Hessian $\nabla_X^2 J$, while also having no guarantee of invertibility far from the critical points of $J$.   
In practice one uses methods like \textsc{l-bfgs} to build an approximation $p_{\text{\textsc{l-bfgs}}}^{(k)} \approx p_{\text N}^{(k)}$ to the search direction at iteration $k$  based on the gradients calculated in iterations  $1,2,\dots,k$ \cite{von2008computational}.  
In passing we note that our numerical experiments suggest that restarting the \textsc{l-bfgs} by erasing the gradient history may improve convergence rates in some situations. 

As mentioned above, in a number of experimentally relevant cases there is also a bound on the maximal and minimal values of the control $u_{\text{min}}\leq u(t) \leq u_{\text{max}}$. These bounds can be accommodated by using a non-linear transformation that makes the control unconstrained as in the Bose-Hubbard example program. An alternative is to add a term in the cost function Eq. (\ref{eq:cost}) that penalizes controls outside the bounds. The latter approach, is known as soft bounds and the QEngine supports a parabolic cost penalty
\begin{equation}
J_{\text{b}}=\frac{\sigma}{2} \int_0^T \Theta(u_{\text{min}}-u)(u-u_{\text{min}})^2 + \Theta(u-u_{\text{max}})(u-u_{\text{max}})^2 \text{d}t, \label{eq:softBounds}
\end{equation}
where $\Theta$ is the Heaviside step function. The weight factor $\sigma$ is typically of the order $\sigma \sim 10^3 - 10^4$ to heavily penalize controls outside the bounds. It is straightforward to calculate the $L^2$ gradient of this term which is
\begin{equation}
\nabla_{L^2} J_{\text{b}} = \sigma \Bigl(\Theta(u_{\text{min}}-u)(u-u_{\text{min}})+\Theta(u-u_{\text{max}})(u-u_{\text{max}})\Bigr).
\end{equation}
This gradient is added to Eq. (\ref{eq:L2}) steering the optimization towards a region inside the bounds.

Collectively the updates using gradients described in Eqs. \eqref{eq:L2}-\eqref{eq:H1} are known as \textsc{grape} algorithms (\textsc{g}radient \textsc{a}scent \textsc{p}ulse \textsc{e}ngineering) \cite{khaneja2005optimal}. 
Numerically solving Eq. \eqref{eq:optimality1} requires discretizing time in steps of $\Delta t$ with a total number of steps $N=\lfloor T/\Delta t\rfloor + 1$. 
In the \textsc{grape} algorithm family, the dimensionality $M$ of the optimization problem is equal to the number of simulation time steps, $M=N$, which is usually on the order of thousands. 

\subsection{\textsc{group}} 
The \textsc{group} algorithm (\textsc{gr}adient \textsc{o}ptimization \textsc{u}sing \textsc{p}arametrization) \cite{sorensen2018group} consists in expanding the control function $u$ on a reduced basis of functions 
$f_m(t)$ where $1 \leq m \leq M$
\begin{align}
u(t;\vec c) = u_0(t) + S(t)\sum_{m=1}^{M} c_m f_m(t) \label{eq:uexpansion},
\end{align}
and performing gradient-based optimization in the $M$-dimensional space of real coefficients  $\vec c = \left[c_1,c_2,\dots,c_M\right]^T$, which is usually on the order of tens, $M\ll N$ \cite{sorensen2018group}. 
This gives \textsc{group} a much smaller optimization dimensionality than \textsc{grape} and is also independent of the duration and size of the time steps. 
In Eq. \eqref{eq:uexpansion} $u_0(t)$ is a reference control and $S(t)$ is any shape-function that goes to zero for $t=0$ and $t=T$, together enforcing appropriate boundary conditions Eq. \eqref{eq:optimality3}.  
Since we are now optimizing the expansion coefficients $\vec c$, the first G\^ ateaux variations of $J$ wrt. $\{\psi(t),\chi(t),\vec c\}$ are set to zero
\begin{align}
D_{\delta\psi}J = D_{\delta\chi}J = D_{\delta \vec c}J= 0.
\end{align}
This produces the same equations of motion as Eqs. \eqref{eq:optimality1}-\eqref{eq:optimality2}. Effectively now $J=J[\vec c]$.
Choosing the inner product to be the usual vector dot product for $X=\mathbb R^M$, the corresponding gradient of $J$ wrt. $\vec c$ is then defined as the element $\vec\nabla_{{\mathbb R}^M}J$ fulfilling the relation
\begin{align}
D_{\delta \vec{c}} J = \langle \vec \nabla_{{\mathbb R}^M} J, \vec{\delta c} \rangle_{{\mathbb R}^M} = \vec\nabla_{{\mathbb R}^M} J \cdot \vec{\delta c}
= \sum_{m=1}^M \frac{\partial J}{\partial c_m} \delta c_m,
\end{align}
where $\delta \vec c$ is an arbitrary variation. This definition of the gradient coincides with the usual definition of the gradient as a column vector of partial derivatives.
The \textsc{group} gradient elements become
\begin{align}
\frac{\partial J}{\partial c_m} = \int_{0}^{T}\bigg(- \Re\Braket{\chi | \frac{d\hat H_0(u)}{du} | \psi} -\gamma \ddot{u} \bigg) S(t)f_m(t)dt  
= \int_0^T \nabla_{L^2}J(t) S(t)f_m(t) dt \label{eq:groupgrad},
\end{align}
where we identified the term in parenthesis to simply be the $L^2$ \textsc{grape} gradient from Eq. \eqref{eq:L2}. Calculating the \textsc{group}
gradient amounts to first calculating the usual $L^2$ \textsc{grape} gradient and subsequently performing $M$ inexpensive one-dimensional integrals  \cite{sorensen2018group}. The coefficients are then iterated according to
\begin{align}
\vec c^{\;(k+1)} = \vec c^{\;(k)} + \alpha^{(k)} \vec p^{\;(k)},
\label{eq:groupupdate}
\end{align}
where $ \vec p^{\;(k)}$ is either the steepest descent direction or the \textsc{l-bfgs} direction both utilizing the gradient Eq. \eqref{eq:groupgrad}.
\subsection{Dressed GROUP}
A caveat of the parametrization Eq. \eqref{eq:uexpansion} is that we may induce local trap minima not inherent to the control problem, but rather due to the parametrization itself. These types of traps are known as artifical traps \cite{rach2015dressing}. A method to escape such traps was proposed in Ref. \cite{rach2015dressing}. We may let $f_m = f_m(t;\vec \theta_m)$ where $\vec \theta_m$ is a set of values that is usually drawn at random. 
For example we may take $f_m(t,\theta_m) = \sin\left((m+\theta_m) \pi t/T\right)$ where $-0.5 \leq \theta_m \leq 0.5$ is drawn from a uniform distribution.
Then, if the algorithm gets trapped at (possibly) an artificial minimum, we set $u_0(t) \leftarrow u(t)$,  re-initialize the algorithm with coefficients $\vec{c} =\vec 0$, and 
draw a new set of basis functions $f_m$ defined by a new set of values $\vec \theta_m$,
\begin{align}
u_0(t) \leftarrow u(t), \quad \vec c \leftarrow 0, \quad  f_m(t; \vec \theta_m) \leftarrow  f_m(t; \vec \theta^*_m),
\label{eq:dressing}
\end{align}
where $\vec \theta^*_m$ are new random values. 
This changes the topology of the optimization landscape and the artificial trap may have been eliminated. 
Effectively, this corresponds to restarting the \textsc{group} algorithm with a new parametrization basis. These restarts are known as superiterations \cite{rach2015dressing}. 
This modification is refered to as dressed \textsc{group}, or d\textsc{group} for short \cite{sorensen2018group}.

\section{Optimal Control Example Programs} \label{sec:qocPrograms}
All algorithms described in the previous section are readily available in the QEngine. These algorithms can be instantiated by the simple API listed in Table \ref{tbl:algs}. 
Additional options such as stopping criteria, data collection, and step size finding method can be set independently of the physical model.
If these are not specified, default options are used.
As a result, the code in the following sections is valid for all example programs. 

We now extend the example programs in section \ref{sec:simulation} by performing optimal control on the systems. Initially we will use the simplest API to perform QOC on the Bose-Hubbard example program. Afterwards we will show a more advanced API for the GPE example program.

\begin{table}[t]
	\begin{center}
		\begin{tabular}{l l}
			\hline
			\multicolumn{1}{c}{\textsc{grape}} & \multicolumn{1}{c}{\textsc{group}} \\
			\hline
			\c{makeGrape\_steepest\_L2(problem)} &  \c{makeGroup\_steepest(problem,basis)}\\ 
			\c{makeGrape\_steepest\_H1(problem)} & \c{makeGroup\_bfgs(problem,basis)}  \\
			\c{makeGrape\_bfgs\_L2(problem)	} & \c{makeDGroup\_steepest(problem,basisMaker)}\\
			\c{makeGrape\_bfgs\_H1(problem)} & \c{makeDGroup\_bfgs(problem,basisMaker)}\\
			\hline
		\end{tabular}
		\caption{Simplest API to instantiate different control algorithms in the QEngine. The \textsc{grape} algorithms (left) only require a state transfer problem object, whereas the \textsc{group} algorithms (right) additionally require a basis specification. 
		The \c{basisMaker} object creates a new basis in each superiteration according to prescription \eqref{eq:dressing}. 
	}
			\label{tbl:algs}
	\end{center}
\end{table}

\subsection{Control of Bose-Hubbard Program -- Simple API}
Here we continue the example program from section \ref{subsec:bosehubbard}. 
In this example we prepare a state transfer problem and solve it using \textsc{grape}, \textsc{group}, and d\textsc{group}.
A state transfer problem is encapsulated by a \c{problem} object,
\begin{lstlisting}
auto problem = makeStateTransferProblem(H, psi_0, psi_t, u, krylovOrder);
\end{lstlisting}
Here \c{H} is the Hamiltonian, \c{psi\_0} is the initial state, \c{psi\_t} is the target state, and \c{u} is the initial control field, which in this case is an exponential ramp.  The \c{krylovOrder} is the order used internally in the timestepper. 
This \c{problem} object maintains a control field and is responsible for calculating the corresponding cost functional and gradient. These quantities are used internally by the optimization algorithm to update the control according to Eq. \eqref{eq:grapeupdate} or \eqref{eq:groupupdate}. 
Having set up the problem, we can apply the different algorithms using the simple API listed in Table \ref{tbl:algs},
\begin{lstlisting}
// GRAPE
auto GRAPE = makeGrape_bfgs_L2(problem);
GRAPE.optimize();
const auto u_grape = GRAPE.problem().control();

// GROUP
const auto M = 60; // basis size
const auto shapeFunction = makeSigmoidShapeFunction(ts,0.999);
const auto maxRand = 0.0; // -maxRand < theta_m < maxRand
const auto basis =  shapeFunction*makeSineBasis(M,u.metaData(),maxRand);

auto GROUP = makeGroup_bfgs(problem,basis);
GROUP.optimize(); // begin optimization
const auto u_group = GROUP.problem().control(); // extract GROUP optimized control

// dGROUP
auto basisMaker = makeBasisMaker([M,maxRand,&u, &shapeFunction]()
{
		return shapeFunction*makeSineBasis(M, u.metaData(), maxRand);
});

auto dGROUP = makeDGroup_bfgs(problem,basisMaker);
dGROUP.optimize(); // begin optimization
const auto u_dgroup = dGROUP.problem().control(); // extract dGROUP optimized control
\end{lstlisting}
After construction, calling \c{.optimize()} begins the optimization algorithm. Once the optimization is completed, the optimized control fields are extracted by \c{.problem().control()}.
\textsc{grape} optimizes the control field directly whereas \textsc{group} uses a reduced basis, which must be supplied by the user -- see Table \ref{tbl:algs}. \textsc{group} also uses a shape function to enforce boundary conditions on the control field  Eq. \eqref{eq:uexpansion}. The default shape function \c{makeShapeFunction} is a symmetric sigmoid function depending on parameters like \c{dt} and the number of controls, which are conveniently supplied through \c{u.metaData()}. 
\textsc{group} uses the same basis for the entire optimization whereas d\textsc{group} uses a new basis in each superiteration through the presecription \eqref{eq:dressing}. The new basis is constructed from the \c{basisMaker} object. 

The result of the d\textsc{group} optimization is shown in Fig. \ref{fig:bhAfter}\textbf{a} where the on-site density is plotted as a function of time for the optimized control. After the vertical dashed line the density and the control is constant since the target state is an eigenstate. In Fig. \ref{fig:bhAfter} the controls from the other optimization algorithms are plotted. In this example the algorithms find very  similar solutions.

\begin{figure}[t]
\begin{minipage}{0.48\linewidth}
	\centering
	\includegraphics[width=1\linewidth]{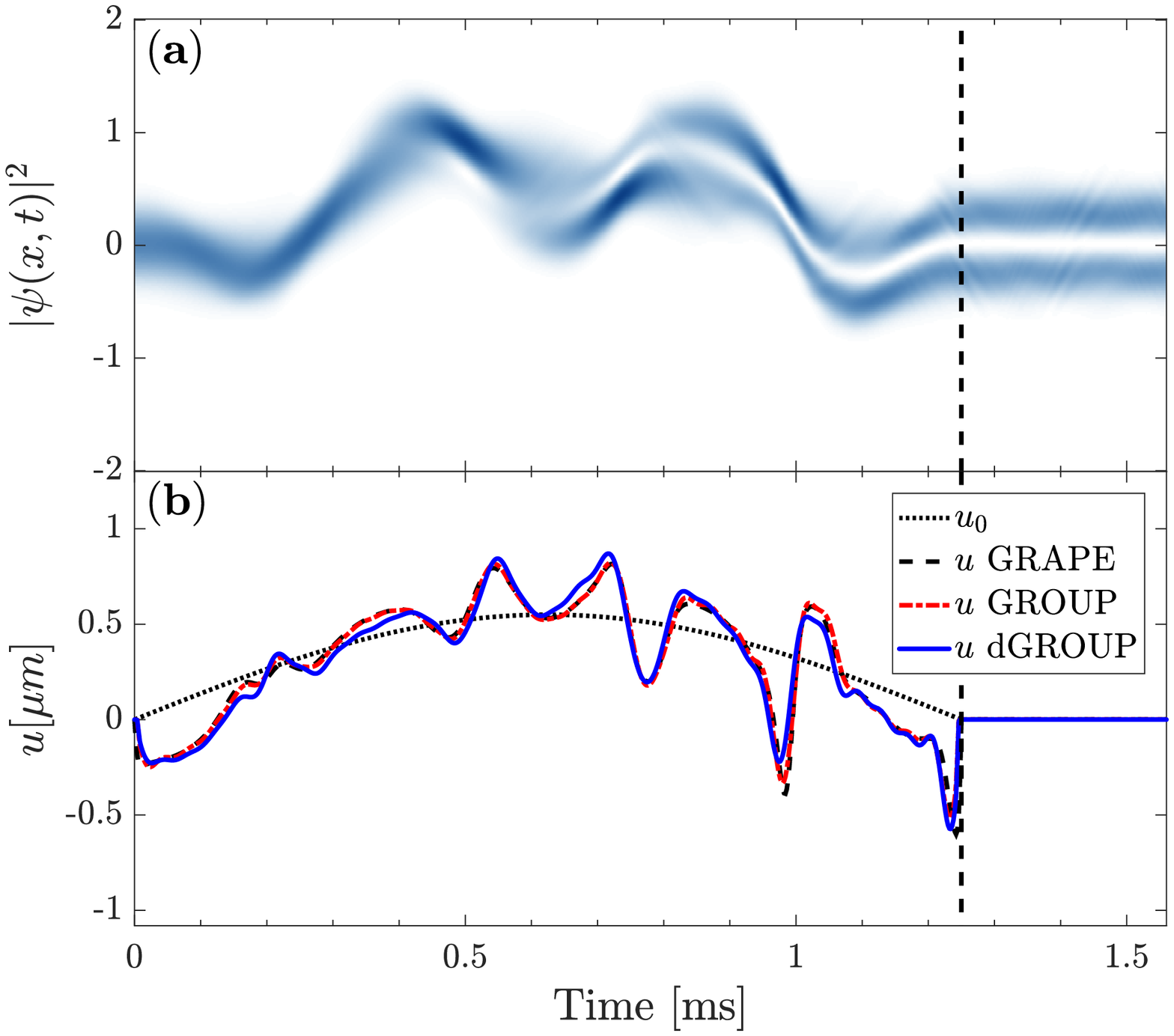}
		\caption{(\textbf{a}) the density of the condensate $|\psi(x,t)|^2$ when propagated along the optimized d\textsc{group} control from the Gross-Pitaevskii example program. (\textbf{b}) the initial control $u_0$ and the optimized controls from \textsc{grape} $F=0.992$, \textsc{group} $F=0.992$ and d\textsc{group} $F=0.994$ -- see legend.}
		\label{fig:gpAfter}
\end{minipage}
\quad
\begin{minipage}{0.48\linewidth}
		\centering
	\includegraphics[width=1.07\linewidth]{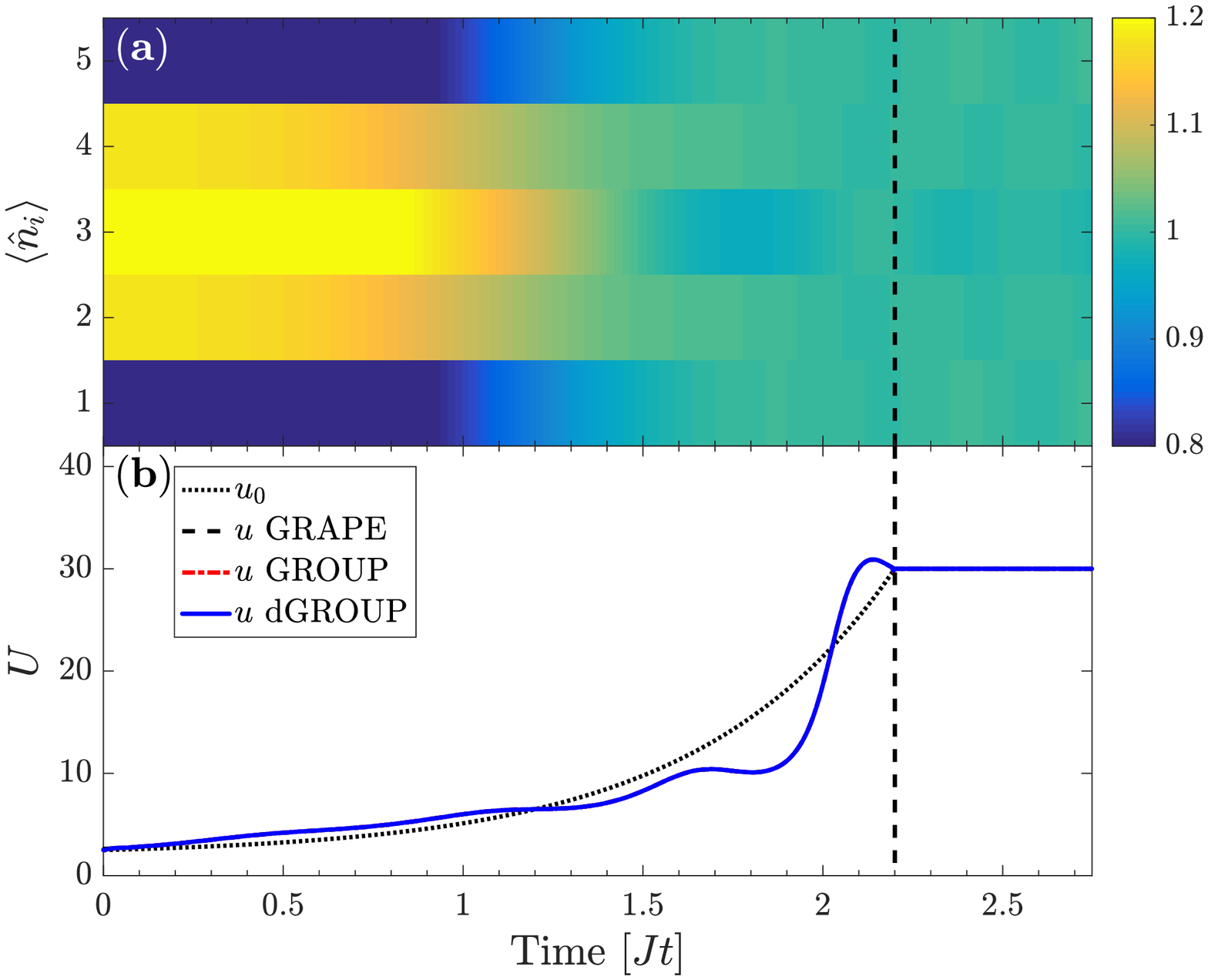}
	\caption{(\textbf{a}) the on-site density $\braket{\hat n_i}$ when propagated along the optimized d\textsc{group} control from the Bose-Hubbard example program. (\textbf{b}) the initial control $u_0$ and the optimized controls from \textsc{grape}, \textsc{group} and d\textsc{group} -- see legend. All algorithms have fidelity $F=0.99$ and find almost identical controls.}
	\label{fig:bhAfter}
\end{minipage}
\end{figure}

\subsection{Control of Gross-Pitaevskii Program -- Advanced API}
In this section we continue the example program from section \ref{subsec:gpeex}. Here we perform \textsc{grape}, \textsc{group}, and d\textsc{group} 
optimizations using the more advanced API. 

Calculating the gradient Eq. \eqref{eq:L2} requires the derivative of the Hamiltonian wrt. the control. The default behavior in the simple API is to calculate the derivate numerically using finite differences. It is more efficient and accurate to manually supply the analytic derivative
\begin{lstlisting}
const auto dHdu_func = [&x,p2,p4,p6](const real u)
{
	auto x_u = x-u;
	auto x_u_Pow2 = x_u*x_u;
	auto x_u_Pow3 = x_u*x_u_Pow2;
	auto x_u_Pow5 = x_u_Pow2*x_u_Pow3;
	
	return  -(2*p2*x_u + 4*p4*x_u_Pow3 + 6*p6*x_u_Pow5);
};

const auto dHdu = makeAnalyticDiffPotential(makePotentialFunction(dHdu_func,u_initial));

auto problem = makeStateTransferProblem(H,dHdu,psi_0,psi_t,u) 
			 + 1e-5*Regularization(u) 
			 + 2e3*Boundaries(u,RVec{-1},RVec{+1});
\end{lstlisting}
Note the resemblance to the cost functional Eq. \eqref{eq:cost} when initializing the state transfer problem. The last term adds soft bounds to the optimization problem as in Eq. \eqref{eq:softBounds} that penalizes control values outside $|u(t)|\geq 1\mu m$, which is set by experimental constraints on the control problem \cite{van2016optimal}.

The QEngine allows for arbitrary stopping conditions within its optimization algorithms. After each iteration the optimizer calls a \c{stopper} object to check if the optimization should be stopped. This object is instantiated by the \c{makeStopper} function. Any callable taking the optimizer type as its argument and returning a boolean can be passed in to this function.
In this example we pass an inline lambda into \c{makeStopper} using \c{auto\&} to deduce the optimizer type. 
\begin{lstlisting}
// Stopper object
const auto stopper = makeStopper([](auto& optimizer) -> bool
{
	bool stop = false;
	if (optimizer.problem().fidelity() > 0.999) 
		{ std::cout << "Fidelity criterion satisfied" << std::endl; stop = true; }
	if (optimizer.previousStepSize() < 1e-7)    
		{ std::cout << "Step size too small" << std::endl; stop = true; };
	if (optimizer.iteration() == 2000) 
		{ std::cout << "Max iterations exceeded" << std::endl; stop = true; }
	if (stop) 
		{std::cout << "STOPPING" << std::endl;}
	return stop;
});
\end{lstlisting}
After each iteration the optimizer also calls a \c{collector} object and defines what should be displayed, saved, and so on in each iteration. 
The \c{collector} works similarly to the \c{stopper} and is instantiated by the \c{makeCollector} function. 
In this example the fidelity of the current control is saved to the \c{DataContainer}, and a status message is printed to console.
\begin{lstlisting}
// Collector object
const auto collector = makeCollector([&dc,n_steps](auto& optimizer) {
	dc["fidelityHistory"].append(optimizer.problem().fidelity());
	std::cout <<
		"ITER "       << optimizer.iteration() << " | " <<
		"fidelity : " << optimizer.problem().fidelity() << "\t " <<
		"stepsize : " << optimizer.stepSize()   << "\t " <<
		"fpp : "      << round(optimizer.problem().nPropagationSteps()/n_steps) << "\t " <<
	std::endl;
});

\end{lstlisting}
Next we define a line search algorithm that calculates a suitable step size $\alpha^{(k)}$ along the search direction $p^{(k)}$ in Eqs. \eqref{eq:grapeupdate} and \eqref{eq:groupupdate}. 
The QEngine supplies an interpolating line search algorithm that can be used out of the box by calling the \c{makeInterpolatingStepSizeFinder} function.  
It is also possible to create custom line search algorithms. As a simple example, a constant step size strategy is commented out to show the interface for custom algorithms. 
\begin{lstlisting}
// Stepsize finder object
const auto maxStepSize = 5.0;
const auto maxInitGuess = 1.0;
const auto stepSizeFinder = makeInterpolatingStepSizeFinder(maxStepSize,maxInitGuess);

// const auto constStepSize=makeStepSizeFinder([](auto& dir,auto& problem,auto& optimizer)
//{
//	return 0.01;
//});
\end{lstlisting}
Restarting the \textsc{l-bfgs} algorithm is sometimes beneficial as noted in section \ref{sec:qoc}. In this example we simply use the default option. 
The optimizer of choice is then created by calling the corresponding \c{make} function with the additional objects defined above.
\begin{lstlisting}
/// GRAPE
auto GRAPE = makeGrape_bfgs_L2(problem,stopper,collector,stepSizeFinder);
collector(GRAPE); // collect iteration 0
GRAPE.optimize(); // begin optimization
const auto u_grape = GRAPE.problem().control(); // extract GRAPE optimized control

/// GROUP
const auto M = 60; // basis size
auto maxRand = 0.0; // -maxRand < theta_m < maxRand
const auto shapeFunction = makeSigmoidShapeFunction(ts,0.999);
const auto basis = shapeFunction*makeSineBasis(basisSize,u.metaData(),maxRand);

problem.update(0*u); // update problem such that reference control will be u_0(t)=0
auto GROUP = makeGroup_bfgs(problem,basis,stopper,collector,stepSizeFinder);

// the initial guess is the first element in the basis, so we may set explicitly:
auto cs = GROUP.problem().coefficients();
cs.at(0).at(0) = initialAmplitude;
GROUP.problem().update(cs); // set initial coefficient vector \vec c = [0.55,0,0,...,0]

collector(GROUP);
GROUP.optimize(); // begin optimization
const auto u_group = GROUP.problem().control(); // extract GROUP optimized control
\end{lstlisting}
To perform the optimization and retrieve the optimized control we invoke \c{.optimize()} on each optimizer. 
The optimization runs until the \c{stopper} function returns true. For the \textsc{group} optimizations we use a sine basis Eq. \eqref{eq:uexpansion} with no randomization. On construction the reference control and coefficients in Eq. \eqref{eq:uexpansion} are set to \c{u} and $\vec c = [0,0,0,\dots,0]^T$. We can manually update them to e.g. $\vec c = [0.55,0,0,\dots,0]^T$ where $0.55$ is the initial amplitude as illustrated. 

For d\textsc{group} we first prepare a shape-function and the \c{BasisMaker} as in the Bose Hubbard example program.  In d\textsc{group} there is also the possibility to supply a user-defined \c{dressedRestarter} object that in each iteration checks if the algorithm should re-initialize with a new random basis as in Eq. \eqref{eq:dressing}, which requires a non-zero bound \c{maxRand} on the random values $\vec \theta_m$. As a simple example we re-initialize the algorithm every 100 iterations or if there is only a small decrease in the cost

\begin{lstlisting}
/// dGROUP
maxRand = 0.1;
const auto basisMaker = makeRandSineBasisMaker(basisSize, shapeFunction, maxRand);

auto restart_func = [tol{ 1e-6 }](const auto& dGROUP) mutable
{
	auto stepSize = dGROUP.stepSize();
	if (stepSize < tol)
	{
		std::cout << "New superiteration in dGROUP algorithm." << std::endl;
		return true;
	}
	return false;
};
const auto dRestarter = makeDressedRestarter(restart_func);
problem.update(0*u);
auto dGROUP = makeDGroup_bfgs(problem, basisMaker, stopper, collector, stepSizeFinder, dRestarter);

dGROUP.problem().update(cs); // set initial coefficient vector \vec c = [0.55,0,0,...,0]

collector(dGROUP);  // collect iteration 0
dGROUP.optimize(); // begin optimization

const auto u_dgroup = dGROUP.problem().control(); // extract dGROUP optimized control
\end{lstlisting}

Finally the data from all the optimizations is saved into a \c{json} file by calling \c{dc.save("gpe-example.json")} or alternatively a \c{mat} file with \c{dc.save("gpe-example.mat")} if matio has been configured.

The result of the d\textsc{group} optimization is shown in Fig. \ref{fig:gpAfter}\textbf{a} where the condensate density is plotted as a function of time. The control and the density is constant after the vertical line since the target state is an eigenstate. The final controls from all the optimization algorithms are displayed in Fig. \ref{fig:gpAfter}\textbf{b}.

\section{Summary and Outlook} \label{sec:summaryOutlook}
We have introduced and described how to use the QEngine. In the example programs we showed how the \c{auto}-syntax combined with factory-functions allows the user to straightforwardly set up optimal control simulations of ultracold atomic systems. However, the QEngine is limited in the number of physical models it supports e.g. we only simulate the dynamics of one-dimensional systems in the current version. We plan to release future versions of the QEngine that can simulate a wider range of dynamics. We also plan to include more sophisticated optimal control for instance the quality of the gradients can be improved by also taking the temporal discretization into account.

\section{Acknowledgments}
This work has funded by the European Research Council and the Lundbeck Foundation. We would also like to thank Birk Skyum for help with testing and building of the QEngine.

\appendix

\section{Units and Nondimensionalization}
\label{ap:Units}
The SI-unit system usually results in very small numerical values for quantum mechanical simulations, which makes simulations impractical or infeasible. For this reason it is beneficial to rescale physical quantities into characteristic scales of the problem such that most values are of order unity. This is achieved using a process known as nondimensionalization where quantities in SI-units are written in product form e.g. $a$ becomes $a=\alpha \tilde{a}$ where $\alpha$ carries both the dimension of $a$ and a magnitude
while $\tilde a$ is a non-dimensional scaling value. This is done for all quantities and substituted into the equations of motion, which leaves a new set of working equations involving only the dimensionless scaling values. As an example consider the GPE example program discussed in section \ref{subsec:gpeex} where the relevant quantities are
\begin{align}
x = \chi \tilde x, \qquad t=\tau \tilde t,\qquad  V=\epsilon \tilde V, \qquad \psi = \xi \tilde\psi, \qquad g_{\text{1D}}  = \gamma \tilde g_{\text{1D}}
\end{align}
We may a priori take length to be measured in micrometer and time to be measured in milliseconds. 
\begin{align}
\chi = 1\mu m,\qquad \tau=1ms
\end{align}
The three remaining units are chosen conveniently as
 \begin{align}
\epsilon = \frac{\hbar^2}{2\kappa m\chi^2}, \qquad \xi = \sqrt\frac{N}{\chi},\qquad \gamma = \frac{\epsilon}{\xi^2}, 
 \end{align}
 where $\kappa$ is the kinetic factor. Substituting into the GPE
\begin{equation}
i\frac{\hbar}{\tau}\xi\frac{\partial\tilde \psi(\tilde x,\tilde t)}{\partial\tilde t} =
 \left(-\kappa \left(\frac{1}{\kappa}\frac{\hbar^2}{2m \chi^2}\right) \frac{\partial^2}{\partial \tilde x^2} + \epsilon\tilde V(\tilde x) + \gamma \xi^2 \tilde g_{\text{1D}} |\tilde \psi(\tilde x,\tilde t)|^2\right) \xi\tilde \psi(\tilde x,\tilde t)
 \end{equation}
Dividing by $\epsilon$ and requring $\hbar/\tau \epsilon=1$ or equivalently $\kappa = \tau \hbar/2 m \chi^2$ gives
\begin{equation}
i \frac{\partial {\tilde \psi}(\tilde x,\tilde t)}{\partial \tilde t} =
 \left(-\kappa \frac{\partial^2}{\partial \tilde x^2} + \tilde V(\tilde x)+ \tilde g_{\text{1D}} |\tilde \psi(\tilde x,\tilde t)|^2\right) \tilde \psi(\tilde x,\tilde t),
\end{equation}
which is the dimensionless form the GPE solved in the example programs. For the potential we find $\tilde p_i$ by comparing
\begin{align}
\tilde V = \frac{V}{\epsilon} = \sum_{i=2,4,6}\frac{p_i}{\epsilon}(x-u)^i  = \sum_{i=2,4,6}\left(\frac{p_i \chi^i}{\epsilon}\right) (\tilde x-\tilde u)^i = \sum_{i=2,4,6} \tilde p_i (\tilde x-\tilde u)^i 
\end{align}
In these units the nondimensionalized scaling values used in the simulation for $N=700$ are
\begin{align}
\kappa = 0.36537,\qquad  \tilde p_2 = 65.8392, \qquad \tilde p_4 = 97.6349,\qquad  \tilde p_6 = -15.3850, \qquad \tilde g_{\text{1D}} = 1.8299.
\end{align}




\bibliographystyle{elsarticle-num}
\bibliography{references}

\begin{thebibliography}{10}
\expandafter\ifx\csname url\endcsname\relax
  \def\url#1{\texttt{#1}}\fi
\expandafter\ifx\csname urlprefix\endcsname\relax\def\urlprefix{URL }\fi
\expandafter\ifx\csname href\endcsname\relax
  \def\href#1#2{#2} \def\path#1{#1}\fi

\bibitem{rosi2014precision}
G.~Rosi, F.~Sorrentino, L.~Cacciapuoti, M.~Prevedelli, G.~Tino, Precision
  measurement of the newtonian gravitational constant using cold atoms, Nature
  510~(7506) (2014) 518.

\bibitem{poli2011precision}
N.~Poli, F.-Y. Wang, M.~Tarallo, A.~Alberti, M.~Prevedelli, G.~Tino, Precision
  measurement of gravity with cold atoms in an optical lattice and comparison
  with a classical gravimeter, Physical review letters 106~(3) (2011) 038501.

\bibitem{schumm2005matter}
T.~Schumm, S.~Hofferberth, L.~M. Andersson, S.~Wildermuth, S.~Groth,
  I.~Bar-Joseph, J.~Schmiedmayer, P.~Kr{\"u}ger, Matter-wave interferometry in
  a double well on an atom chip, Nature physics 1~(1) (2005) 57--62.

\bibitem{andersson2002multimode}
E.~Andersson, T.~Calarco, R.~Folman, M.~Andersson, B.~Hessmo, J.~Schmiedmayer,
  Multimode interferometer for guided matter waves, Physical review letters
  88~(10) (2002) 100401.

\bibitem{wang2005atom}
Y.-J. Wang, D.~Z. Anderson, V.~M. Bright, E.~A. Cornell, Q.~Diot, T.~Kishimoto,
  M.~Prentiss, R.~Saravanan, S.~R. Segal, S.~Wu, Atom michelson interferometer
  on a chip using a bose-einstein condensate, Physical review letters 94~(9)
  (2005) 090405.

\bibitem{bloch2012quantum}
I.~Bloch, J.~Dalibard, S.~Nascimbene, Quantum simulations with ultracold
  quantum gases, Nature Physics 8~(4) (2012) 267--276.

\bibitem{georgescu2014quantum}
I.~Georgescu, S.~Ashhab, F.~Nori, Quantum simulation, Reviews of Modern Physics
  86~(1) (2014) 153.

\bibitem{kaufman2015entangling}
A.~Kaufman, B.~Lester, M.~Foss-Feig, M.~Wall, A.~Rey, C.~Regal, Entangling two
  transportable neutral atoms via local spin exchange, Nature 527~(7577) (2015)
  208.

\bibitem{de2008optimal}
G.~De~Chiara, T.~Calarco, M.~Anderlini, S.~Montangero, P.~Lee, B.~Brown,
  W.~Phillips, J.~Porto, Optimal control of atom transport for quantum gates in
  optical lattices, Physical Review A 77~(5) (2008) 052333.

\bibitem{weitenberg2011quantum}
C.~Weitenberg, S.~Kuhr, K.~M{\o}lmer, J.~F. Sherson, Quantum computation
  architecture using optical tweezers, Physical Review A 84~(3) (2011) 032322.

\bibitem{bloch2008many}
I.~Bloch, J.~Dalibard, W.~Zwerger, Many-body physics with ultracold gases,
  Reviews of modern physics 80~(3) (2008) 885.

\bibitem{van2016optimal}
S.~van Frank, M.~Bonneau, J.~Schmiedmayer, S.~Hild, C.~Gross, M.~Cheneau,
  I.~Bloch, T.~Pichler, A.~Negretti, T.~Calarco, et~al., Optimal control of
  complex atomic quantum systems, Scientific reports 6.

\bibitem{glaser2015training}
S.~J. Glaser, U.~Boscain, T.~Calarco, C.~P. Koch, W.~K{\"o}ckenberger,
  R.~Kosloff, I.~Kuprov, B.~Luy, S.~Schirmer, T.~Schulte-Herbr{\"u}ggen,
  et~al., Training schr{\"o}dinger's cat: quantum optimal control, The European
  Physical Journal D 69~(12) (2015) 279.

\bibitem{werschnik2007quantum}
J.~Werschnik, E.~Gross, Quantum optimal control theory, Journal of Physics B:
  Atomic, Molecular and Optical Physics 40~(18) (2007) R175.

\bibitem{jager2013optimal}
G.~J{\"a}ger, U.~Hohenester, Optimal quantum control of bose-einstein
  condensates in magnetic microtraps: Consideration of filter effects, Physical
  Review A 88~(3) (2013) 035601.

\bibitem{jager2014optimal}
G.~J{\"a}ger, D.~M. Reich, M.~H. Goerz, C.~P. Koch, U.~Hohenester, Optimal
  quantum control of bose-einstein condensates in magnetic microtraps:
  Comparison of gradient-ascent-pulse-engineering and krotov optimization
  schemes, Physical Review A 90~(3) (2014) 033628.

\bibitem{ndong2010vibrational}
M.~Ndong, C.~P. Koch, Vibrational stabilization of ultracold krb molecules: A
  comparative study, Physical Review A 82~(4) (2010) 043437.

\bibitem{koch2004stabilization}
C.~P. Koch, J.~P. Palao, R.~Kosloff, F.~Masnou-Seeuws, Stabilization of
  ultracold molecules using optimal control theory, Physical Review A 70~(1)
  (2004) 013402.

\bibitem{doria2011optimal}
P.~Doria, T.~Calarco, S.~Montangero, Optimal control technique for many-body
  quantum dynamics, Physical review letters 106~(19) (2011) 190501.

\bibitem{bucker2013vibrational}
R.~B{\"u}cker, T.~Berrada, S.~Van~Frank, J.-F. Schaff, T.~Schumm,
  J.~Schmiedmayer, G.~J{\"a}ger, J.~Grond, U.~Hohenester, Vibrational state
  inversion of a bose--einstein condensate: optimal control and state
  tomography, Journal of Physics B: Atomic, Molecular and Optical Physics
  46~(10) (2013) 104012.

\bibitem{caneva2009optimal}
T.~Caneva, M.~Murphy, T.~Calarco, R.~Fazio, S.~Montangero, V.~Giovannetti,
  G.~E. Santoro, Optimal control at the quantum speed limit, Physical review
  letters 103~(24) (2009) 240501.

\bibitem{hegerfeldt2013driving}
G.~C. Hegerfeldt, Driving at the quantum speed limit: optimal control of a
  two-level system, Physical review letters 111~(26) (2013) 260501.

\bibitem{deffner2017quantum}
S.~Deffner, S.~Campbell, Quantum speed limits: from heisenberg’s uncertainty
  principle to optimal quantum control, Journal of Physics A: Mathematical and
  Theoretical 50~(45) (2017) 453001.

\bibitem{brouzos2015quantum}
I.~Brouzos, A.~I. Streltsov, A.~Negretti, R.~S. Said, T.~Caneva, S.~Montangero,
  T.~Calarco, Quantum speed limit and optimal control of many-boson dynamics,
  Physical Review A 92~(6) (2015) 062110.

\bibitem{khaneja2005optimal}
N.~Khaneja, T.~Reiss, C.~Kehlet, T.~Schulte-Herbr{\"u}ggen, S.~J. Glaser,
  Optimal control of coupled spin dynamics: design of nmr pulse sequences by
  gradient ascent algorithms, Journal of magnetic resonance 172~(2) (2005)
  296--305.

\bibitem{assion1998control}
A.~Assion, T.~Baumert, M.~Bergt, T.~Brixner, B.~Kiefer, V.~Seyfried,
  M.~Strehle, G.~Gerber, Control of chemical reactions by feedback-optimized
  phase-shaped femtosecond laser pulses, Science 282~(5390) (1998) 919--922.

\bibitem{dolde2014high}
F.~Dolde, V.~Bergholm, Y.~Wang, I.~Jakobi, B.~Naydenov, S.~Pezzagna, J.~Meijer,
  F.~Jelezko, P.~Neumann, T.~Schulte-Herbr{\"u}ggen, et~al., High-fidelity spin
  entanglement using optimal control, Nature communications 5 (2014) 3371.

\bibitem{mennemann2015optimal}
J.-F. Mennemann, D.~Matthes, R.-M. Weish{\"a}upl, T.~Langen, Optimal control of
  bose--einstein condensates in three dimensions, New Journal of Physics
  17~(11) (2015) 113027.

\bibitem{hohenester2007optimal}
U.~Hohenester, P.~K. Rekdal, A.~Borz{\`\i}, J.~Schmiedmayer, Optimal quantum
  control of bose-einstein condensates in magnetic microtraps, Physical Review
  A 75~(2) (2007) 023602.

\bibitem{hohenester2014octbec}
U.~Hohenester, Octbec—a matlab toolbox for optimal quantum control of
  bose--einstein condensates, Computer Physics Communications 185~(1) (2014)
  194--216.

\bibitem{machnes2011comparing}
S.~Machnes, U.~Sander, S.~Glaser, P.~de~Fouquieres, A.~Gruslys, S.~Schirmer,
  T.~Schulte-Herbr{\"u}ggen, Comparing, optimizing, and benchmarking
  quantum-control algorithms in a unifying programming framework, Physical
  Review A 84~(2) (2011) 022305.

\bibitem{schmidt2017wavepacket1}
B.~Schmidt, U.~Lorenz, Wavepacket: A matlab package for numerical quantum
  dynamics. i: Closed quantum systems and discrete variable representations,
  Computer Physics Communications 213 (2017) 223--234.

\bibitem{schmidt2018wavepacket}
B.~Schmidt, C.~Hartmann, Wavepacket: A matlab package for numerical quantum
  dynamics. ii: Open quantum systems, optimal control, and model reduction,
  Computer Physics Communications 228 (2018) 229--244.

\bibitem{johansson2012qutip}
J.~Johansson, P.~Nation, F.~Nori, Qutip: An open-source python framework for
  the dynamics of open quantum systems, Computer Physics Communications 183~(8)
  (2012) 1760--1772.

\bibitem{armadillo}
C.~Sanderson, R.~Curtin, Armadillo: a template-based c++ library for linear
  algebra, Journal of Open Source Software 1 (2016) 26.

\bibitem{sorensen2018group}
J.~J. S{\o}rensen, M.~Aranburu, T.~Heinzel, J.~Sherson, Gradient based quantum
  optimal control in a chopped basis, arXiv preprint arXiv:1802.07509.

\bibitem{dalfovo1999theory}
F.~Dalfovo, S.~Giorgini, L.~P. Pitaevskii, S.~Stringari, Theory of
  bose-einstein condensation in trapped gases, Reviews of Modern Physics 71~(3)
  (1999) 463.

\bibitem{olshanii}
M.~Olshanii, Atomic scattering in the presence of an external confinement and a
  gas of impenetrable bosons, Physical Review Letters 81~(5) (1998) 938.

\bibitem{dion2007ground}
C.~M. Dion, E.~Canc{\`e}s, Ground state of the time-independent
  gross--pitaevskii equation, Computer physics communications 177~(10) (2007)
  787--798.

\bibitem{taha1984analytical}
T.~R. Taha, M.~I. Ablowitz, Analytical and numerical aspects of certain
  nonlinear evolution equations. ii. numerical, nonlinear schr{\"o}dinger
  equation, Journal of Computational Physics 55~(2) (1984) 203--230.

\bibitem{park1986unitary}
T.~J. Park, J.~Light, Unitary quantum time evolution by iterative lanczos
  reduction, The Journal of chemical physics 85~(10) (1986) 5870--5876.

\bibitem{barredo2016atom}
D.~Barredo, S.~De~L{\'e}s{\'e}leuc, V.~Lienhard, T.~Lahaye, A.~Browaeys, An
  atom-by-atom assembler of defect-free arbitrary two-dimensional atomic
  arrays, Science 354~(6315) (2016) 1021--1023.

\bibitem{endres2016atom}
M.~Endres, H.~Bernien, A.~Keesling, H.~Levine, E.~R. Anschuetz, A.~Krajenbrink,
  C.~Senko, V.~Vuletic, M.~Greiner, M.~D. Lukin, Atom-by-atom assembly of
  defect-free one-dimensional cold atom arrays, Science 354~(6315) (2016)
  1024--1027.

\bibitem{deChiaraGate}
G.~De~Chiara, T.~Calarco, M.~Anderlini, S.~Montangero, P.~Lee, B.~Brown,
  W.~Phillips, J.~Porto, Optimal control of atom transport for quantum gates in
  optical lattices, Physical Review A 77~(5) (2008) 052333.

\bibitem{anderlini2007controlled}
M.~Anderlini, P.~J. Lee, B.~L. Brown, J.~Sebby-Strabley, W.~D. Phillips,
  J.~Porto, Controlled exchange interaction between pairs of neutral atoms in
  an optical lattice, Nature 448~(7152) (2007) 452--456.

\bibitem{bucker2011twin}
R.~B{\"u}cker, J.~Grond, S.~Manz, T.~Berrada, T.~Betz, C.~Koller,
  U.~Hohenester, T.~Schumm, A.~Perrin, J.~Schmiedmayer, Twin-atom beams, Nature
  Physics 7~(8) (2011) 608--611.

\bibitem{gerbier2004quasi}
F.~Gerbier, Quasi-1d bose-einstein condensates in the dimensional crossover
  regime, EPL (Europhysics Letters) 66~(6) (2004) 771.

\bibitem{two_qubit_quantum_gate_by_cold_controlled_collisons}
D.~Jaksch, H.-J. Briegel, J.~Cirac, C.~Gardiner, P.~Zoller, Entanglement of
  atoms via cold controlled collisions, Physical Review Letters 82~(9) (1999)
  1975.

\bibitem{quantum_gate_BH_via_collisions}
O.~Mandel, M.~Greiner, A.~Widera, T.~Rom, T.~W. H{\"a}nsch, I.~Bloch,
  Controlled collisions for multi-particle entanglement of optically trapped
  atoms, Nature 425~(6961) (2003) 937.

\bibitem{laser_induced_quantum_gate_operations}
G.~K. Brennen, C.~M. Caves, P.~S. Jessen, I.~H. Deutsch, Quantum logic gates in
  optical lattices, Physical Review Letters 82~(5) (1999) 1060.

\bibitem{FastQuantumGates}
D.~Jaksch, J.~Cirac, P.~Zoller, S.~Rolston, R.~C{\^o}t{\'e}, M.~Lukin, Fast
  quantum gates for neutral atoms, Physical Review Letters 85~(10) (2000) 2208.

\bibitem{Toffili_gate_1D_lattice}
J.~K. Pachos, P.~L. Knight, Quantum computation with a one-dimensional optical
  lattice, Physical review letters 91~(10) (2003) 107902.

\bibitem{quantum_simulator}
E.~Jan{\'e}, G.~Vidal, W.~D{\"u}r, P.~Zoller, J.~I. Cirac, Simulation of
  quantum dynamics with quantum optical systems, Quantum Information \&
  Computation 3~(1) (2003) 15--37.

\bibitem{single_atom_transistor}
A.~Micheli, A.~Daley, D.~Jaksch, P.~Zoller, Single atom transistor in a 1d
  optical lattice, Physical review letters 93~(14) (2004) 140408.

\bibitem{braun2015emergence}
S.~Braun, M.~Friesdorf, S.~S. Hodgman, M.~Schreiber, J.~P. Ronzheimer,
  A.~Riera, M.~del Rey, I.~Bloch, J.~Eisert, U.~Schneider, Emergence of
  coherence and the dynamics of quantum phase transitions, Proceedings of the
  National Academy of Sciences 112~(12) (2015) 3641--3646.

\bibitem{BH_experimental_realization}
M.~Greiner, O.~Mandel, T.~Esslinger, T.~W. H{\"a}nsch, I.~Bloch, Quantum phase
  transition from a superfluid to a mott insulator in a gas of ultracold atoms,
  nature 415~(6867) (2002) 39.

\bibitem{ramping_optimization_zakrzewski}
J.~Zakrzewski, D.~Delande, Breakdown of adiabaticity when loading ultracold
  atoms in optical lattices, Physical Review A 80~(1) (2009) 013602.

\bibitem{rosi2013fast}
S.~Rosi, A.~Bernard, N.~Fabbri, L.~Fallani, C.~Fort, M.~Inguscio, T.~Calarco,
  S.~Montangero, Fast closed-loop optimal control of ultracold atoms in an
  optical lattice, Physical Review A 88~(2) (2013) 021601.

\bibitem{von2008computational}
G.~Von~Winckel, A.~Borz{\`\i}, Computational techniques for a quantum control
  problem with h1-cost, Inverse Problems 24~(3) (2008) 034007.

\bibitem{nocedal2006numerical}
J.~Nocedal, S.~J. Wright, Numerical optimization 2nd (2006).

\bibitem{rach2015dressing}
N.~Rach, M.~M. M{\"u}ller, T.~Calarco, S.~Montangero, Dressing the
  chopped-random-basis optimization: A bandwidth-limited access to the
  trap-free landscape, Physical Review A 92~(6) (2015) 062343.

\end{thebibliography}







\end{small}

\end{document}